\documentclass[a4paper,11pt]{article}
\usepackage{pos}
\usepackage{microtype}
\DeclareMathOperator{\Arg}{Arg}
\def\app#1#2{%
  \mathrel{%
    \setbox0=\hbox{$#1\sim$}%
    \setbox2=\hbox{%
      \rlap{\hbox{$#1\propto$}}%
      \lower1.1\ht0\box0%
    }%
    \raise0.25\ht2\box2%
  }%
}
\def\approxprop{\mathpalette\app\relax}

\title{Neural Network Field Transformation\\and Its Application in HMC}
\ShortTitle{Neural Network Field Transformation and Its Application in HMC}

\author*[a,b]{Xiao-Yong Jin}

\affiliation[a]{Argonne Leadership Computing Facility, Argonne National Laboratory,\\
	Lemont, IL 60439, USA}
\affiliation[b]{Computational Science Division, Argonne National Laboratory,\\
	Lemont, IL 60439, USA}

\emailAdd{xjin@anl.gov}

\abstract{%
	We propose a generic construction of Lie group agnostic
	and gauge covariant neural networks,
	and introduce constraints to make the neural networks
	continuous differentiable and invertible.
	We combine such neural networks and build gauge field
	transformations that is suitable for Hybrid Monte Carlo (HMC).
	We use HMC to sample lattice gauge configurations in the transformed
	space by the neural network parameterized gauge field transformations.
	Tested with 2D U(1) pure gauge systems at a range of couplings and
	lattice sizes,
	compared with direct HMC sampling,
	the neural network transformed HMC (NTHMC) generates Markov chains
	of gauge configurations with improved tunneling of topological charges,
	while allowing less force calculations as the lattice coupling increases.
}

\FullConference{%
 The 38th International Symposium on Lattice Field Theory, LATTICE2021
  26th-30th July, 2021
  Zoom/Gather@Massachusetts Institute of Technology
}

\begin{document}
\maketitle

\section{\label{sec:intro}Introduction}

When properly trained,
neural networks have the potential of approximating arbitrary
functions with finite computational cost.
By design typical neural networks work on numerical tensors that
live on a flat space of real domain,
such as the common representations of images, sound waves, and natural languages.
There are ongoing active development on curved space with known curvature.
For gauge theories of our interests,
we need neural network layers that work on
the degree of freedoms themselves from compact Lie groups
on the lattice links,
which constitute the dynamics of our theory~\cite{Boyda:2020hsi,Tomiya:2021ywc}.

One application of neural network function approximation is field transformation.
With a transformation approximating a trivializing map,
Hybrid Monte Carlo (HMC)~\cite{Duane:1987de}
can work in the approximately trivialized lattice fields
and generate gauge configurations with less autocorrelation~\cite{Luscher:2009eq}.

We propose a generic construction of Lie group agnostic
and gauge covariant neural network architecture
as fundamental building blocks,
in section~\ref{sec:nnft}.
Section~\ref{sec:nthmc} introduces some constraints to the neural networks,
which allow us to create a continuous differentiable and invertible gauge field transformation
suitable for use with HMC.
As a practical example,
section~\ref{sec:2du1} details our experiment using HMC with the neural network
parameterized field transformation (NTHMC) on two dimensional U(1) lattice gauge field.
We conclude in section~\ref{sec:concl}.

\section{\label{sec:nnft}Gauge Covariant Neural Networks and Field Transformation}

We build the elementary neural network field transformation layer
by extending gauge covariant
update to the links of values in a Lie group,
\begin{equation}
	\label{eq:transform}
	U_{x,\mu}\rightarrow \tilde{U}_{x,\mu} = e^{\Pi_{x,\mu}} U_{x,\mu},
\end{equation}
where $U_{x,\mu}$ denotes the gauge link from lattice site $x$ to $x+\hat{\mu}$,
in the direction $\mu$,
and the exponent,
\begin{equation}
	\label{eq:def-sum-wl}
	\Pi_{x,\mu} = \sum_l \epsilon_{x,\mu,l} \partial_{x,\mu} W_l,
\end{equation}
forms from a list of Wilson loops $W_l$ taken a group derivative,
$\partial_{x,\mu}$,
with respect to $U_{x,\mu}$.
Unlike the common lattice field theory applications,
our extension for neural networks lies in the local coefficients,
$\epsilon_{x,\mu,l}$,
as arbitrary scalar valued functions ($\mathcal{N}_l$)
of gauge invariant quantities ($X$, $Y$, \ldots),
\begin{equation}
	\label{eq:def-eps}
	\epsilon_{x,\mu,l} = c_l \tan^{-1}\big[ \mathcal{N}_l(X,Y,\ldots) \big].
\end{equation}
The additional application of $\tan^{-1}$ and the multiplication of scalar
coefficients $c_l$ serve to constrain the possible values of $\epsilon$.
Naturally $\mathcal{N}_l$ can be any neural network.

\section{\label{sec:nthmc}HMC with Neural Network Field Transformation}

Instead of sampling the target field configurations $\{U\}$
according to the action $S(U)$,
we use a continuously differentiable bijective map $\mathcal{F}$
with $U=\mathcal{F}(\tilde{U})$,
and apply a change of variables in the path integral for an observable
$\mathcal{O}$,
similar to reference~\cite{Luscher:2009eq},
\begin{equation}
	\langle \mathcal{O}\rangle =
	\frac{1}{\mathcal{Z}} \int
	\mathrm{D}[U] \mathcal{O}(U) e^{-S(U)} =
	\frac{1}{\mathcal{Z}} \int
	\mathrm{D}[\tilde{U}] \mathcal{O}\big(\mathcal{F}(\tilde{U})\big) e^{-S_{\text{FT}}(\tilde{U})},
\end{equation}
with the effective action after the field transformation,
\begin{equation}
	S_{\text{FT}}(\tilde{U}) =
	S\big(\mathcal{F}(\tilde{U})\big) - \ln\big|\mathcal{F}_*(\tilde{U})\big|,
\end{equation}
and the Jacobian of the transformation,
\begin{equation}
	\mathcal{F}_*(\tilde{U}) = \frac{\partial\mathcal{F}(\tilde{U})}{\partial \tilde{U}}.
\end{equation}
We generate a Markov Chain of field configurations $\{\tilde{U}\}$
using HMC as usual with the action given as $S_{\text{FT}}(\tilde{U})$.
An arbitrary neural network parameterized field transformation
as in equation~\eqref{eq:transform} can be used as $\mathcal{F}$,
as long as the coefficients $c_l$ in equation~\eqref{eq:def-eps}
are in the range such that the Jacobian of the transformation
remain positive definite.

In order to have a tractable Jacobian,
we update the whole lattice field in steps such that each step
only update a subset of the lattice field,
as suggested in reference~\cite{Luscher:2009eq}.
For each step, we choose $W_l$ in equation~\eqref{eq:def-sum-wl}
and $\epsilon_{x,\mu,l}$ in equation~\eqref{eq:def-eps}
independent of the subset gauge links $U_{x,\mu,l}$ under transformation
in equation~\eqref{eq:transform}.
It makes the determination of $c_l$ simple for a positive
definite Jacobian.

For translational and rotational symmetry of the lattice,
as well as scalability,
we choose to use convolutional neural networks (CNN) as $\mathcal{N}_l$
in equation~\eqref{eq:def-eps} with gauge invariant inputs ($X$, $Y$, \ldots)
taken from Wilson loops of different sizes.
This also makes $\epsilon_{x,\mu,l}$ depend on nearby Wilson loops
with the locality controlled by the kernel sizes of the CNN.

\section{\label{sec:2du1}Results in 2D U(1) Lattice Gauge Theory}

As an experiment of the neural networks,
we apply the neural network parameterized field transformation
HMC (NTHMC) for U(1) lattice gauge theory
in two dimensions.
There are methods that improve tunneling of topological charges
while generating Monte Carlo samples of gauge configurations~\cite{Eichhorn:2021ccz}.
Our experiment here aims to see how the trained transformations
perform in NTHMC compared with direct HMC sampling.
Our code is available online~\cite{web:nthmc}, with more data
than described here.

\begin{figure}
\centering                                        
\includegraphics[width=0.618\linewidth]{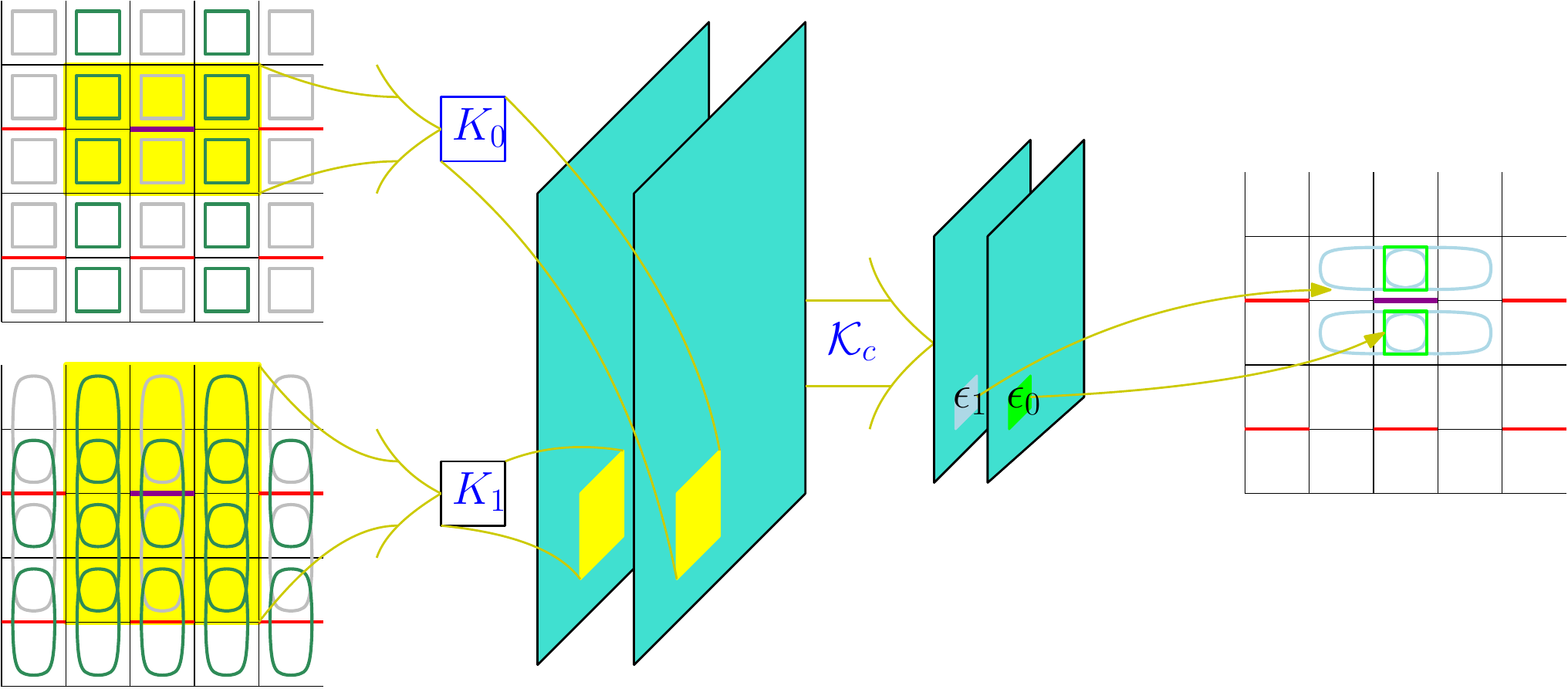}
\caption{\label{fig:filters} Gauge covariant neural network filters for local link update coefficients.}
\end{figure}

We use a fixed neural network architecture as shown in figure~\ref{fig:filters}.
It presents one step that updates the subset of gauge links in red.
Eight such steps on different links update the whole lattice.
There are two series of CNN's that respectively use
plaquette and rectangle Wilson loops as inputs with green color,
which are independent of the red gauge links.
The CNN kernel $\mathcal{K}_0$ and $\mathcal{K}_1$ center around
the dark red links and have a size of the yellow shade.
Another series of CNN with kernel $\mathcal{K}_c$ takes the stacked
results of the previous network and generates output with two channels,
one each with scalar values,
which after applying $\tan^{-1}$ and scaled with $c_l$ as in
equation~\eqref{eq:def-eps} produce the $\epsilon_{x,\mu,l}$
for updating that particular gauge link in dark red with
the derivatives of the plaquette and rectangle loops.

Specifically we use one layer of CNN with four channels
and a kernel size of $3\times 2$ and $3\times 3$ respectively
for $\mathcal{K}_0$ and $\mathcal{K}_1$,
and two layers of CNN with four and two channels
both with a kernel size of $3\times 3$
for $\mathcal{K}_c$,
and all CNN's use Gaussian error linear unit (GELU) activation
function~\cite{DBLP:journals/corr/HendrycksG16} except for the last one.
The entire transformation used in HMC consists of sixteen
individual steps that goes through the gauge links in the entire lattice twice,
with each step using distinct network weights.

We perform simulations using HMC with the Wilson
plaquette gauge action for 2D U(1) gauge field, using a trajectory length
of four molecular dynamics time unit (MDTU),
with Omelyan's second order minimum norm integrator~\cite{OMELYAN2003272} and step size tuned
to have an acceptance rate around $80\%$.
Though after the transformation with the effective action $S_{\text{FT}}(\tilde{U})$,
the MDTU no longer has the same meaning as in direct HMC with original action,
we still keep the same value of the trajectory length.
We compute the topological charge, $Q=\frac{1}{2\pi}\sum_x \Arg P_x$,
on each configurations and compute the autocorrelation,
using the fact that $\langle Q \rangle=0$,
\begin{equation}
	\label{eq:autocorr}
\begin{split}
	\Gamma_t(\delta) \equiv
	\frac{\langle Q_\tau Q_{\tau+\delta} \rangle}
	{\langle Q^2 \rangle}
	&=
	1 -
	\frac{
		\langle Q^2 \rangle - \langle Q_\tau Q_{\tau+\delta} \rangle
	}{
		V \chi_t^\infty(\beta)
	}\\
	&=
	1 -
	\frac{
		\langle (Q_\tau - Q_{\tau+\delta})^2 \rangle
	}{
		2 V \chi_t^\infty(\beta)
	},
\end{split}
\end{equation}
where $V$ is the lattice volume,
$\beta$ is the lattice coupling constant,
$Q_\tau$ and $Q_{\tau+\delta}$ are the topological charge of the configurations
separated by $\delta$ MDTU,
and the infinite volume topological susceptibility~\cite{Bonati:2019ylr} is,
\begin{equation}
	\chi_t^\infty(\beta) = \frac{
		\int_{-\pi}^\pi \left(\frac{\phi}{2\pi}\right)^2 e^{\beta \cos\phi} \mathrm{d}\phi
	}{
		\int_{-\pi}^\pi e^{\beta \cos\phi} \mathrm{d}\phi
	}.
\end{equation}
As $\langle Q^2 \rangle$ and $\langle Q_\tau Q_{\tau+\delta} \rangle$
are highly correlated in finite number of samples,
the subtraction in the right hand side of equation~\eqref{eq:autocorr}
produces less statistical uncertainty than directly computing
the left hand side.

\begin{figure}
\centering                                        
\includegraphics[width=0.49\linewidth]{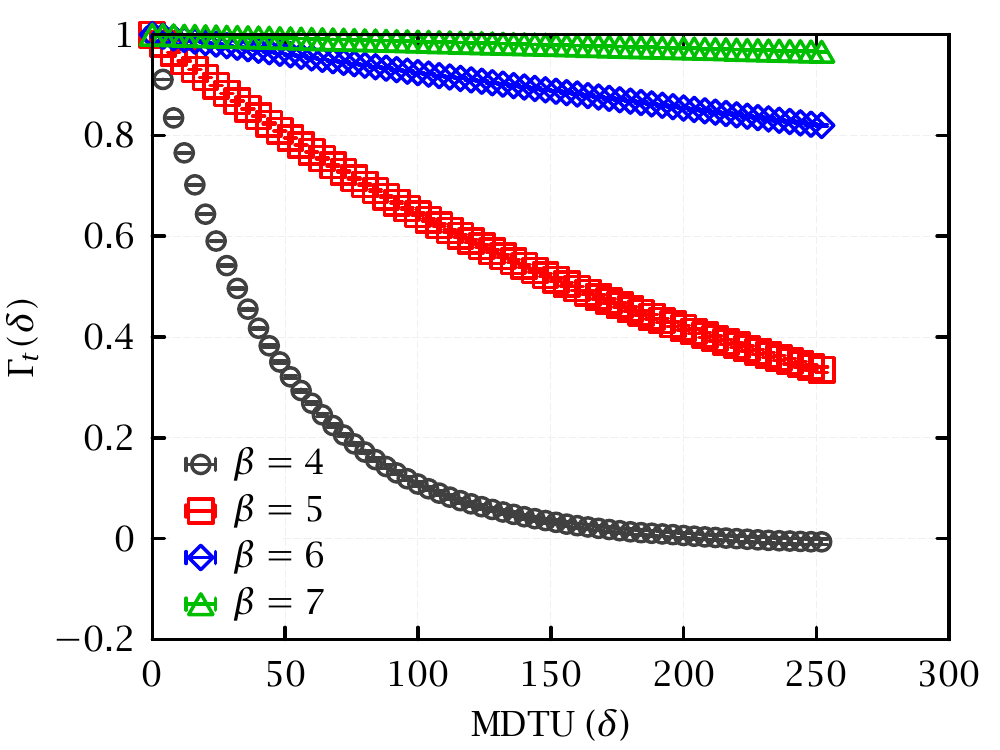}
\includegraphics[width=0.49\linewidth]{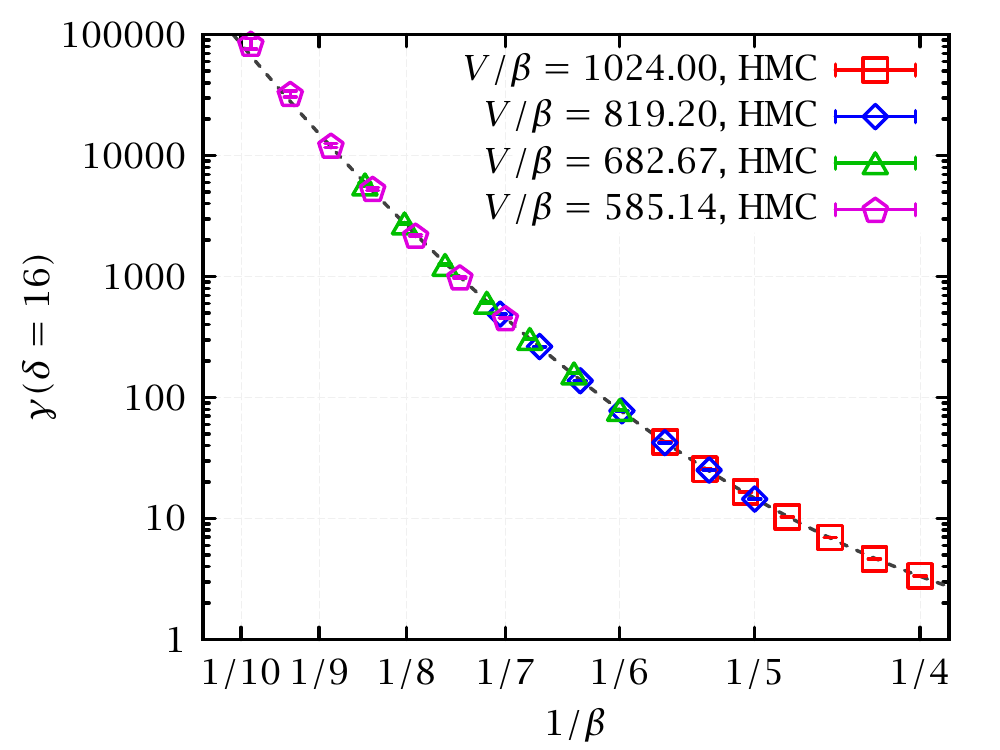}
\caption{\label{fig:autocorr_hmc}
	Left: Topological charge autocorrelation from direct HMC with $64^2$ lattices.
	Right: Power law scaling of $\gamma(\delta=16)$ with HMC
	versus $1/\beta$ with fixed $V/\beta$, or constant physical volumes.}
\end{figure}

Critical slowing down of the tunneling of the topological charge
is evident in the Markov chain generated by direct HMC sampling.
The left panel of figure~\ref{fig:autocorr_hmc} shows the autocorrelation
of the topological charge at $\beta=4$, $5$, $6$, and $7$ with $64^2$ lattices
generated with direct HMC.
While the autocorrelation vanishes at around 250 MDTU for $\beta=4$,
the values for $\beta=7$ barely differs from one.
The right panel shows the power law scaling of $\gamma(\delta=16)$ versus $\beta$,
with
\begin{equation}
	\gamma(\delta) = \frac{1}{1-\Gamma_t(\delta)} \approxprop \tau_{\text{int}}.
\end{equation}
As the inverse of the deviation of the autocorrelation away from one
approximately proportional to the integrated autocorrelation length,
we use $\gamma(\delta=16)$ to estimate the relative independence of
the generated lattice configurations.
The dashed line in the figure is a quadratic fit in the log-log scale to all the points
to guide the eye.

We train the neural network parameterized field transformation by minimizing
the difference between the force of the transformed action
and the force of the original U(1) gauge action on $\tilde{U}$ at a fixed $\beta=2.5$.
Concretely the loss function on a transformed field $\tilde{U}$ is,
\begin{equation}
	\label{eq:loss}
	\mathcal{L}(\beta, \tilde{U}) = \sum_{p\in\{2,4,6,8,10,\infty\}}
	\frac{c_p}{V^{1/p}}
	\Bigg|\Bigg|
		\frac{\partial S_{\text{FT}}(\beta, \tilde{U})}{\partial \tilde{U}} -
		\frac{\partial S(\beta=2.5, \tilde{U})}{\partial \tilde{U}}
	\Bigg|\Bigg|_p,
\end{equation}
where $||\cdot||_p$ denotes $p$-norm,
and $c_p$ controls the optimization to favor volume averages or peaks on individual links.
We set $c_2=c_4=c_6=c_8=c_{\infty}=1$ and $c_{10}=0$ for the models presented here,
unless specified otherwise.
We start from randomized neural networks weights,
train the models from $\beta=3$,
and after that load the trained model and continue training at $\beta=4$.
We repeat this procedure at $\beta=5$ and $6$.
We generate $2^{17}$ independent gauge configurations at each $\beta$ before training.
At each $\beta$ value, the training uses Adam optimizer~\cite{Kingma:2014vow},
and goes through pre-generated $2^{17}$ configurations once,
with a batch size of $128$.
With $64^2$ lattices, the training for each $\beta$ value took about $35$ minutes
on a Tesla V100-SXM2-16GB GPU.

\begin{figure}
\centering                                        
\includegraphics[width=0.735\linewidth]{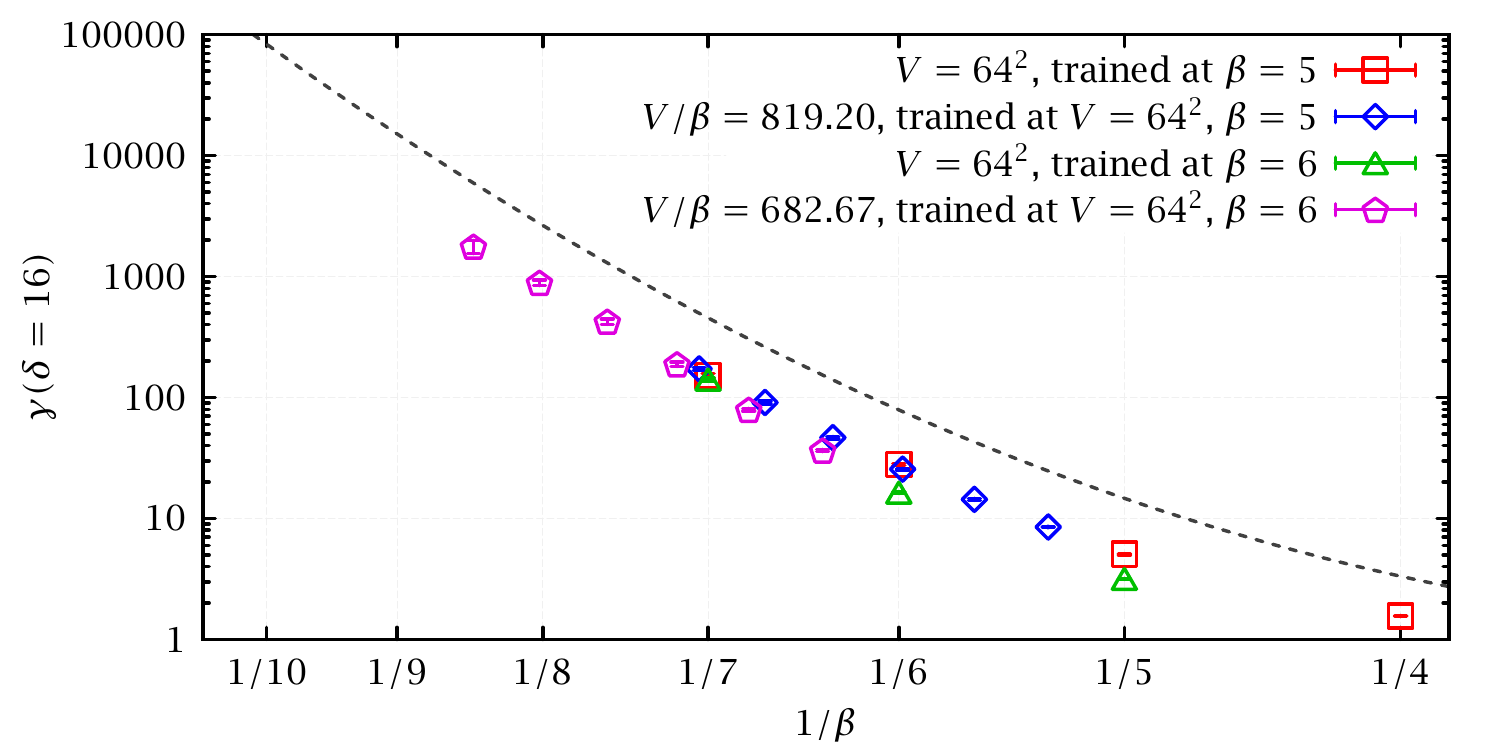}
\caption{\label{fig:autocorr_nthmc}
	Power law scaling of $\gamma(\delta=16)$ with HMC
	using two trained models of neural network parameterized field transformation.
	Dashed line from figure~\ref{fig:autocorr_hmc} indicates direct HMC values.}
\end{figure}

The trained models used as transformations in HMC appear to improve
the tunneling of topological charges in successive Markov Chain states.
Figure~\ref{fig:autocorr_nthmc} shows the same power law scaling
as in the right panel of figure~\ref{fig:autocorr_hmc},
with the dashed line denotes the values from direct HMC without transformation.
The figure contains HMC runs with two different models trained with
a lattice volume of $64^2$ at $\beta=5$ and $6$ respectively.
We employ the models for a fixed volume at $64^2$ at different $\beta$ values,
and for fixed $V/\beta$ values with volumes of $66^2$, $68^2$, $70^2$, $72^2$,
$74^2$, and $76^2$.
It seems that a single model applied to different volumes and $\beta$ values
shows the same scaling coefficients as direct HMC without transformations.
The tunneling improves from the model trained at $\beta=5$ to the model
trained at $\beta=6$.

\begin{figure}
\centering
\includegraphics[width=\linewidth]{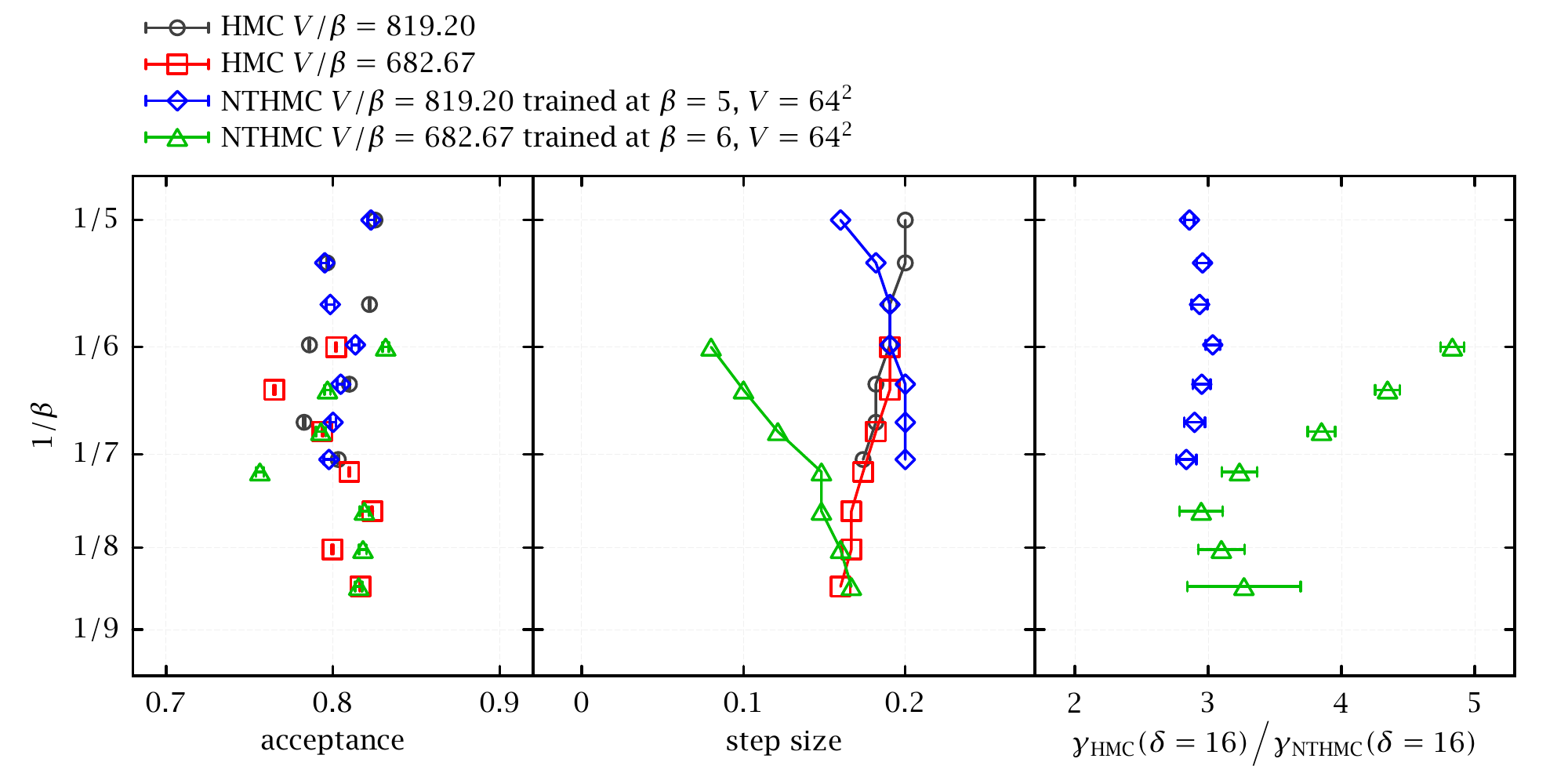}
\caption{\label{fig:nthmc_stepsize_fixBV}
	Acceptance rate, MD step size, and improvement
	in tunneling of topological charges with trained
	models of field transformation with two fixed values of $V/\beta$,
	corresponding to $V=64^2$, $66^2$, $68^2$, $70^2$, $72^2$, $74^2$, and $76^2$.
	NTHMC denotes HMC with neural network
	parameterized field transformation,
	using models trained at $\beta=5$ and $6$ with $V=64^2$.}
\end{figure}

For understanding actual simulation cost,
we study how the Molecular Dynamics (MD) step size changes,
with different models,
as we use a fixed trajectory length of 4,
and tune the step size to have an acceptance rate at around $80\%$.
Figure~\ref{fig:nthmc_stepsize_fixBV} shows the acceptance rate, step size,
and corresponding improvement in autocorrelation of HMC with neural network
parameterized field transformations (NTHMC) against direct HMC,
using the same trained models as in figure~\ref{fig:autocorr_nthmc}.
With acceptance rate around $80\%$, the step sizes required by HMC
reduces with increasing $\beta$,
while the step sizes required by NTHMC increases.
Therefore with the trained models of neural network parameterized
field transformations,
in order to achieve a constant acceptance rate,
we are able to reduce the numbers of force evaluations per trajectory
as the lattice coupling $\beta$ increases.

\begin{figure}
\centering
\includegraphics[width=\linewidth]{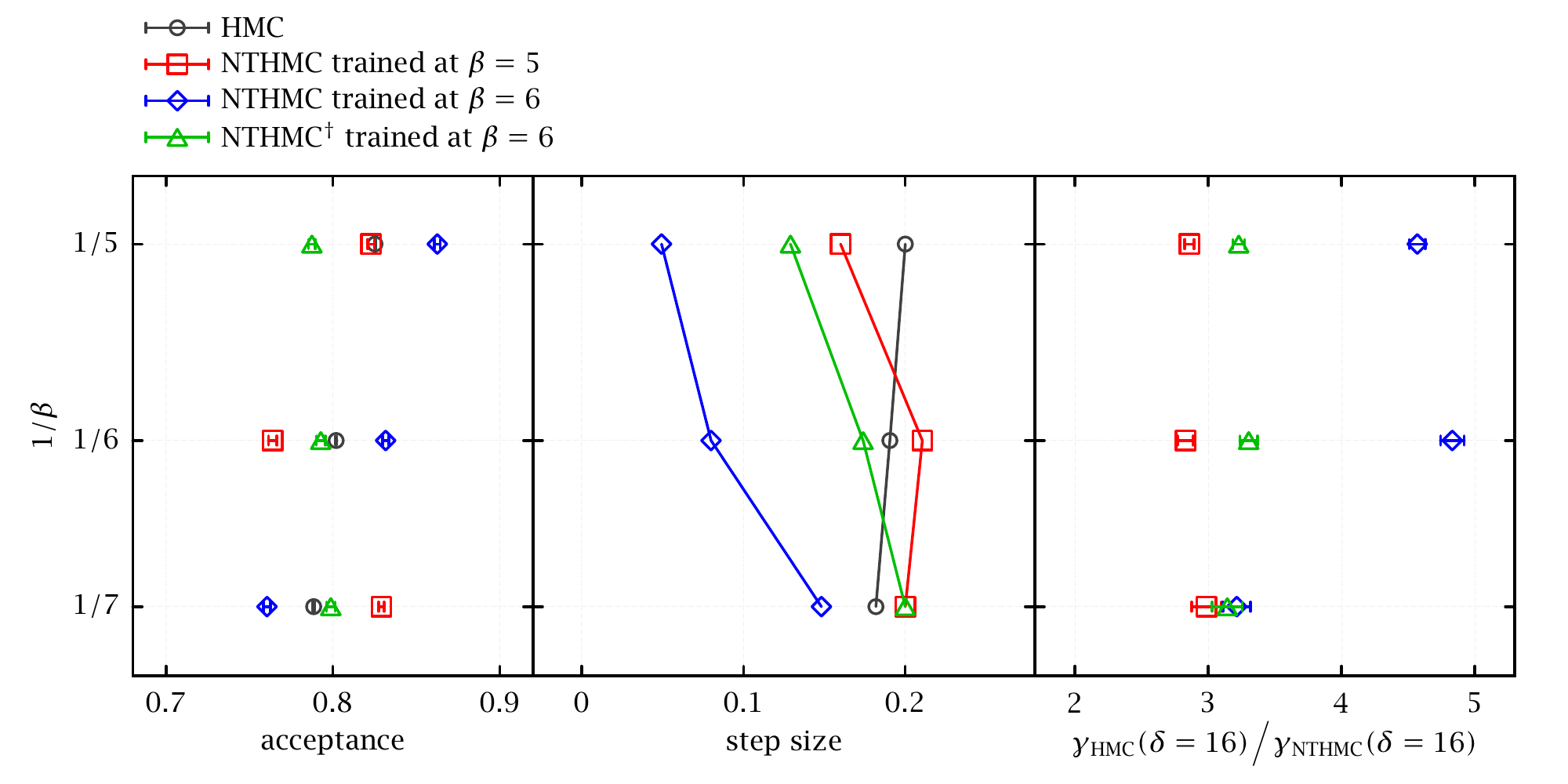}
\caption{\label{fig:nthmc_stepsize_l64}
	Acceptance rate, MD step size, and improvement
	in tunneling of topological charges with trained
	models of field transformation with $64^2$ lattices,
	at $\beta=5$, $6$, and $7$.
	NTHMC denotes HMC with neural network
	parameterized field transformation.  NTHMC$^\dag$ is another model trained
	at $\beta=6$ while setting $c_8=c_{10}=5$ in the loss function,
	equation~\eqref{eq:loss}.}
\end{figure}

We see the similar behavior with a fixed lattice size of $64^2$.
Figure~\ref{fig:nthmc_stepsize_l64} contains two models shown in
figure~\ref{fig:autocorr_nthmc},
and one additional model labeled NTHMC$^\dag$,
which is also from training at $\beta=6$ but with the $8$-norm and $10$-norm
coefficients in the loss function, equation~\eqref{eq:loss}, set to
$c_8=c_{10}=5$.
This attempt at reducing the large peak of the gauge forces results
in less improvement in the tunneling of topological charges,
but significantly increases the allowed MD step size.
Unlike with direct HMC,
both models allow us to use larger step sizes at $\beta=7$ than
the step sizes used at $\beta=6$,
with $\beta=5$ requires the smallest step size.

\section{\label{sec:concl}Conclusion}

We propose a generic construction of Lie group agnostic and gauge covariant
neural networks.
Such construction is a basic building block for designing complex
neural network architectures that work with lattice gauge fields.
We use the proposed construction to
build a neural network architecture for gauge field transformations,
by introducing constraints for a tractable and positive definite Jacobian.
Thus the neural network parameterized
transformation is continuous differentiable and invertible.

We apply the transformation to 2D U(1) pure gauge system and uses HMC to
sample configurations in the transformed field domain.
We train the transformation to match with gauge forces at a stronger coupling.
Using a fixed trajectory length,
HMC with neural network parameterized field transformation (NTHMC) is able to
generate gauge configurations with less autocorrelation in topological charges
for a range of lattice couplings and volumes using a single trained
transformation model
than direct gauge generation with HMC without field transformation.

Our test shows a trade-off between low autocorrelation and large
MD step size.
We are able to make the trained model to favor one or the other by
changing the loss function to use different weights for average gauge force or
maximum gauge force on individual links during training.
In contrast to direct HMC without transformations,
for keeping a relative constant acceptance rate,
the MD step sizes required by NTHMC
with all of the trained transformation models increase with
increasing lattice coupling $\beta$,
allowing less force calculations per trajectory.

\acknowledgments

We would like to thank
Peter Boyle, Norman Christ, Sam Foreman, Taku Izubuchi,
Luchang Jin, Chulwoo Jung, James Osborn, Akio Tomiya,
and other ECP collaborators for insightful discussions and support.
This research was supported by the Exascale Computing Project
(17-SC-20-SC), a collaborative effort of the U.S. Department of
Energy Office of Science and the National Nuclear Security
Administration.
We gratefully acknowledge the computing resources provided and
operated by the Joint Laboratory for System Evaluation (JLSE) at
Argonne National Laboratory.

\bibliographystyle{JHEP}
\bibliography{ref}

\end{document}